\documentclass[trackchanges,twocolumn]{aastex701}

\begin{document}

\title{Stream-Driven Ignition of a Realistic Undisturbed Helium Shell on a White Dwarf}

\author[orcid=0009-0003-0182-2954,gname=Nethra,sname=Rajavel]{Nethra Rajavel}
\affiliation{Department of Physics \& Astronomy, The University of Alabama, Tuscaloosa, AL, USA}
\email[show]{nrajavel@crimson.ua.edu}

\author[orcid=0000-0002-9538-5948, gname=Dean, sname=Townsley]{Dean M. Townsley}
\affiliation{Department of Physics \& Astronomy, The University of Alabama, Tuscaloosa, AL, USA}
\email[]{dmtownsley@ua.edu}

\author[orcid=0000-0002-9632-6106, gname=Ken, sname=Shen]{Ken J. Shen}
\affiliation{Department of Astronomy and Theoretical Astrophysics Center, University of California, Berkeley, CA, USA} 
\email[]{kenshen@astro.berkeley.edu}

\begin{abstract}

The dynamically driven double-degenerate double-detonation (D$^6$) model has emerged as a promising progenitor scenario for Type Ia supernovae. In this model, a carbon-oxygen white dwarf (WD) in a close double WD binary undergoes a double-detonation triggered by dynamical mass transfer from its companion. The mass transfer stream directly impacts the surface of the primary WD, potentially igniting a helium detonation in the surface layer. The resulting shock converges in the core, triggering a carbon detonation that ultimately unbinds the star. While previous studies have demonstrated the viability of this mechanism, the conditions under which the helium shell ignites remain uncertain. We perform two-dimensional simulations using FLASH to study the feasibility of helium detonation ignition driven by stream impact on a carbon-oxygen WD with a realistic, unmixed composition profile left at the end of the previous helium-shell burning phase before the WD was formed. We model WDs with masses of 0.8, 0.9, and 1.0 $\,M_\odot$ and vary stream properties for each case. We find that stream impact leads to helium ignition in all cases, with ignition occurring at varying distances from the impact point. Ignition consistently occurs near the core-shell transition, highlighting the importance of a realistic WD profile. However, the ignition mechanism differs between models and can arise from interactions of neighboring hot, mixed regions rather than a single point. Ultimately, all but one sustain a propagating detonation capable of traversing the WD surface, which is suggestive of the D$^6$ model's viability as a progenitor scenario for Type Ia supernovae.

\end{abstract}



\section{Introduction}  \label{sec:intro}



Type Ia supernovae (SNe Ia), which are generally thought to be thermonuclear explosions of carbon-oxygen (CO) white dwarfs (WDs) in binary systems, have been a matter of study for several decades with multiple theorized progenitor models and explosion mechanisms. SNe Ia are considered ``standardizable candles" that aid in studying dark energy which is why it is even more crucial to understand how they occur (e.g., \citet{Ruiter2025}). 

WDs are categorized as near-Chandrasekhar (near-Ch) or sub-Chandrasekhar mass (sub-Ch) depending on how close the mass is to the Chandrasekhar limit of 1.4 $\,M_\odot$ and the theorized explosion mechanisms for each of these categories differ. Also, the mechanisms differ depending on if the binary is a single degenerate system, meaning a binary system where only one of the stars is a WD, or a double degenerate system, where both stars are WDs. For a near-Ch WD in a single degenerate system, one often-discussed explosion mechanism is the deflagration-to-detonation transition \citep{Khokhlov1997}, but there are also several others (e.g. pure deflagration \citep{Ropke2007} and the gravitationally confined detonation \citep{Plewa2004}). For a near-Ch mass WD in a double degenerate system, the merger of the two WDs, forming a super-Chandrasekhar mass remnant, was considered for many years to be the mechanism that produces type Ia supernovae \citep{Webbink1984}, however, this scenario might instead lead to accretion-induced collapse \citep{Saio1998}. Sub-Ch mass WDs can explode through the double-detonation mechanism \citep{WoosleyWeaver,Livne1991}. According to the double-detonation mechanism, the shock from an initial helium detonation in the shell can converge in the CO core of the WD to cause a secondary detonation that unbinds the entire star. It has been extensively studied over the past decade \citep{Fink2010,Raskin2012, MollWoosley2013, Fenn2016,Garcia-Senz_2018,Tanikawa2019,Townsley_2019,Gronow2020,Gronowetal2021,Boos_2021,Collins2022,Pakmor2022,Roy2022,Shen_2024, Michaelis2025}. The double-detonation mechanism is possible in both single and double degenerate systems. In the D$^6$ model (dynamically driven double degenerate double detonation model), unstable mass transfer from the companion WD, when the WDs are close enough and about to merge, can lead to a detonation of the helium shell on the primary WD and thereafter form a second detonation in the core, causing the entire star to explode \citep{Guillochonetal2010, Dan2011, Pakmor2013, Shen2018b}.
In a single degenerate system, stable mass transfer from a helium-rich companion may trigger the helium detonation necessary for the double-detonation mechanism. Recently, there has been growing support for the double-detonation model, and more specifically, the D$^6$ model as the predominant explosion mechanism \citep{Liu2023}. Although this paragraph summarizes some of the possible explosion mechanisms and progenitor scenarios, for a more complete description, the reader is referred to the relevant reviews cited here \citep{Maoz, Liu2023,Ruiter2025}. 

The double-detonation of the primary sub-Ch mass WD in a double degenerate system (or the D$^6$ model) has proven to be a promising scenario. However, several open questions remain, such as the fate of
the companion WD. The recent discovery of hypervelocity WDs from Gaia DR2 \citep{Shen2018b} and DR3 \citep{El-Badry2023} might answer this question as they are proposed to be the companion WD that has been kicked out from the binary system due to its orbital motion and the rapid removal of the primary's gravitational potential \citep{Tanikawa2018, Bhat2026, Shen2025}. Similarly, the quadruple detonation mechanism, which is the double-detonation of both WDs in a double degenerate system, has been shown to produce spectra comparable to observed SNe Ia and to a scenario where only the primary WD detonates \citep{Pakmor2022,Boos2024, Pollin2024}. Another such question that needs to be answered is how exactly the helium shell detonation is ignited for the double-detonation mechanism. 

Previous studies of the D$^6$ scenario have explored the phase of dynamical mass transfer \citep{Guillochonetal2010, Pakmor2013, Liu2017,Pakmor2022} from the companion to the primary WD just before the WDs are about to merge. The dynamical mass transfer in these simulations ignites a detonation in the helium shell which creates a shock that then converges in the core to detonate the CO core. However, the nature of the ignition deserves in-depth analysis in order to allow an understanding of where and when, during the interaction of the two stars, the detonation will take place.  It is important to characterize the influence of pre-ignition structures in the flowing material and the WD surface. As critical structures are identified, it is similarly necessary to pay careful attention to resolving these structures and processes during their formation and interaction. Additionally, understanding how these details vary across the expected range of scenarios (masses, mass ratios, mass transfer stage) is necessary for robust application of the mode or modes of ignition to astrophysical scenarios for the supernova. Previously, various authors have studied the conditions under which helium detonations ignite and propagate in the shell, including the critical hot spot sizes needed \citep{Holcombe2013}, the effects of CO pollution and nuclear reaction networks on propagation \citep{Moore_2013, Shen_2014}, and whether ignition can arise naturally from accretion through convective instabilities \citep{Glasner2018} or in 1D accretion models \citep{Iwata2022, Kumar2023}.

More recently, \citet{Rajavel2025} modeled two-dimensional simulations local to the surface of the primary WD, with an emphasis on the mass transfer stream and the ignition of the helium detonation. Their key findings are: (1) ignition can happen as the stream is impacting the surface or later once flow over the surface is developed, (2) the hot material deposited by the interaction of the stream with the surface can be important in constraining the stream in a way that later leads to ignition, and (3) the density of the layer is mostly important in determining whether a detonation can propagate once it is ignited, rather than constraining the local ignition itself. While these simulations were successful in producing and characterizing the shell detonation, the WD surface structure used had a boundary between the surface He-rich layer and the interior carbon-oxygen layer that was just a local discontinuity. This is unrealistic since the remnant composition profile between the shell and core left behind during the previous helium shell burning phase before the WD formed should be a smooth, resolved structure at the resolution used in the simulations. Additionally, the ignition seen in \citet{Rajavel2025} occurred at the shell/core interface implying that some amount of carbon was essential for a helium detonation as seen by previous studies \citep{Shen_2014}. Studies have also shown that the primary WD must have a thin helium shell so that the ejected material from the helium detonation is not rich in $^{56}$Ni and more generally iron-group elements \citep{Townsley_2019}. However, the helium shells that were simulated and studied in \citet{Rajavel2025} were thick and too dense. Therefore, the cases that ignited and were able to propagate a detonation in their work had higher helium shell base densities than expected. This could again be because the WD did not have a realistic profile and therefore, the ignition required a thicker shell than it would have needed for a realistic profile WD.  

\citet{Shen_2024} studied the propagation of a helium detonation on the surface of different mass WDs. These realistic WD profiles were constructed with the stellar evolution code \texttt{MESA} \citep{Paxton2011, Paxton2013, Paxton2015, Paxton2018, Paxton2019, Jermyn2023} and a hot spot was used to ignite a detonation on the surface of the WD. They observed that the helium detonation was able to propagate on the surface for a helium shell base density much lower than what was observed in \citet{Rajavel2025}. This further demonstrated the necessity of a realistic WD profile. 

This paper is a follow-up study to \citet{Rajavel2025} and mainly focuses on employing the local stream-WD model from that work, but with more realistic WD profiles. In particular, we map the 1D MESA WD profile used in \citet{Shen_2024} onto that WD model. In section \ref{sec:method} we describe the methods and modifications to the model from \citet{Rajavel2025}. In section \ref{sec:results} we present the results. In section \ref{sec:discussion} we discuss the implications of the results, and in section \ref{sec:conclusion} we summarize and conclude the paper.

\section{Methods}  \label{sec:method}

In this paper, which serves as a follow-up to \citet{Rajavel2025}, we use the same setup utilized in that work with modifications to the boundary conditions and the WD profile. In both this study and the previous one, we simulate the interaction of the mass transfer stream with the surface of the primary WD, assuming a plane-parallel approximation for the WD surface. Such a simulation focuses on modeling the helium detonation triggered by the impact of the mass transfer stream from the companion WD. All simulations are performed in two dimensions. As in \citet{Rajavel2025}, we use the FLASH \citep{Fryxell, DUBEY2009512, Dubey2013, Dubey2014} code (version 4.6) to conduct the simulations, coupled with a 55-isotope nuclear reaction network from MESA which contains the following isotopes: neutrons, $^{1}$H, $^{4}$He, $^{11}$B, $^{12-13}$C, $^{13-15}$N, $^{15-17}$O, $^{18}$F, $^{19-22}$Ne, $^{22-23}$Na, $^{23-26}$Mg, $^{25-27}$Al, $^{28-30}$Si, $^{29-31}$P, $^{31-33}$S, $^{33-35}$Cl, $^{36-39}$Ar, $^{39}$K, $^{40}$Ca, $^{43}$Sc, $^{44}$Ti, $^{47}$V, $^{48}$Cr, $^{51}$Mn, $^{52,56}$Fe, $^{55}$Co, and $^{56,58-59}$Ni. The right and left boundaries use periodic boundary conditions so that the flow of material in the model domain mimics the flow of material around a spherical surface. The bottom boundary of the domain uses a hydrostatic boundary condition \citep{Zingale2002} where the normal velocity is handled by a reflective boundary condition. For a more detailed description of the simulation setup, refer to \citet{Rajavel2025}. The only modifications to the previous setup are the WD profile and the top boundary condition of the domain which will be discussed in the following subsections.

In our simulations, we use adaptive mesh refinement with refinement triggered by variations in density, $^{4}$He mass fraction, temperature and $^{22}$Ne mass fraction. The largest cell sizes are square and $\approx 78\ $ km in each dimension, and refinement is allowed to decrease the cell size (by splitting cells in
half in each dimension) to a minimum cell size of $\approx 10\ $km.

\subsection{Realistic WD composition profile}

\citet{Shen_2024} generated realistic WD profiles using MESA for WDs ranging from 0.5 to 1.1 $\,M_\odot$ and mapped these profiles into FLASH to perform 2D simulations. In that work, the authors placed a hot spot in the transition layer, the region between the core and the shell, to investigate whether the hot spot could produce a detonation capable of propagating across the WD surface and whether the resulting helium detonation could trigger a secondary carbon detonation. Their results showed that most WDs were able to sustain a detonation in the transition layer, except for the 1.0 $\,M_\odot$ and 1.1 $\,M_\odot$ cases, which required an additional layer of accreted material from the companion to support a detonation.

This motivated the present work to investigate whether the mass transfer stream can ignite a detonation in a WD with a realistic profile. To explore this, we adapted our existing model by incorporating the realistic WD profiles from \citet{Shen_2024} into our setup. Specifically, we selected a reference point within the transition layer where $^{12}$C and $^{4}$He have the same abundance and mapped the abundance, temperature and density both upward into the shell and downward into the core. All elements with mass fractions greater than $10^{-6}$ in the WD profile from MESA are mapped onto our WD. The mass fraction of $^{16}$O was set as the difference between 1 and the sum of the mass fraction of the other elements so that all the mass fractions add to 1. Keeping the temperature constant in each of the cells, the density was modified at each of these points based on hydrostatic equilibrium calculations. It was then verified that the resulting profile was stable during several seconds of hydrodynamic evolution using only the vertical direction (1D) and without any stream.

\begin{figure}[ht!]
\centering
\includegraphics[width=0.5\textwidth]{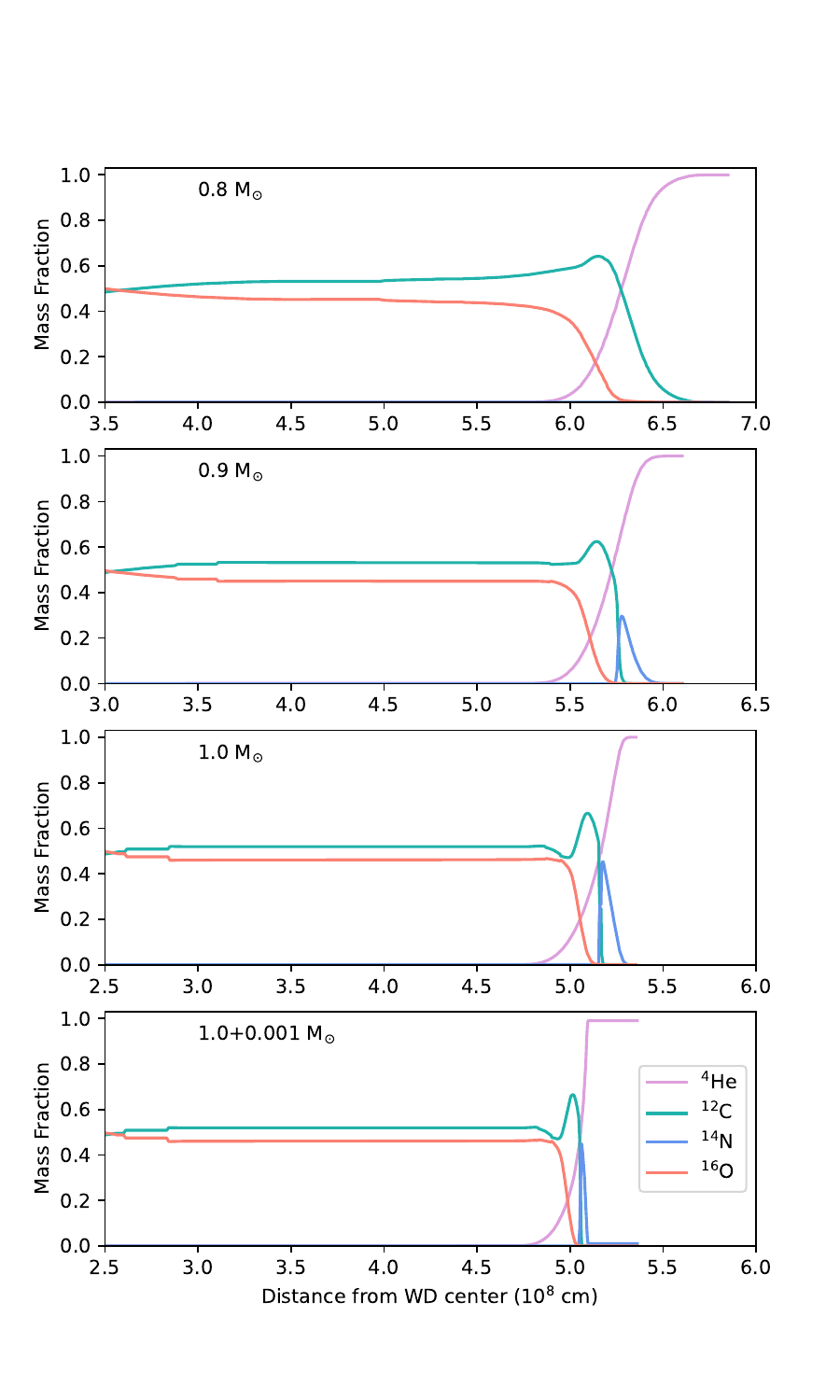}
\caption{1D abundance profiles of different mass WDs taken from \citet{Shen_2024} that are used in this work. Note that only the region inside the domain of our models is shown here. For clarity, only the most abundant species are displayed.}
\label{fig:profile}
\end{figure}

One thing to note about our approach is that, due to hydrostatic equilibrium, the region just outside the WD (hereafter referred to as fluff) has a higher temperature ($10^8$\ K) than from the profile (around 25,000\ K). Additionally, the density profile obtained under hydrostatic equilibrium increasingly diverges from the 1D WD profile toward the interior of the star. This is due to the difference in geometry, and particularly that the gravitational acceleration is uniform in our plane parallel setup, while it varies with radius in the spherical model. Since we are choosing to keep the composition profiles fixed in radius, the surface zone also is at a fixed radius. The resulting hydrostatic equilibrium profile in the plane parallel, uniform gravity geometry has a lower density at the transition layer, being 0.86, 0.75 and 0.92 of the density at the reference point
in the source profile respectively for the 0.8, 1.0 and 1.0+0.001 $\,M_\odot$ cases. The abundance profiles are depicted in Figure \ref{fig:profile}. Since we do not need to simulate the center of the star in detail, we restrict the computational domain so that it begins around halfway into the core. Specifically, this is at a height of $2.5 \times 10^8$\ cm from the center for the 1.0 $\,M_\odot$ WD, $3 \times 10^8$\ cm for the 0.9 $\,M_\odot$ WD and $3.5 \times 10^8$\ cm for the 0.8 $\,M_\odot$ WD. 


\begin{figure}[ht!]
\centering
\includegraphics[width=0.5\textwidth]{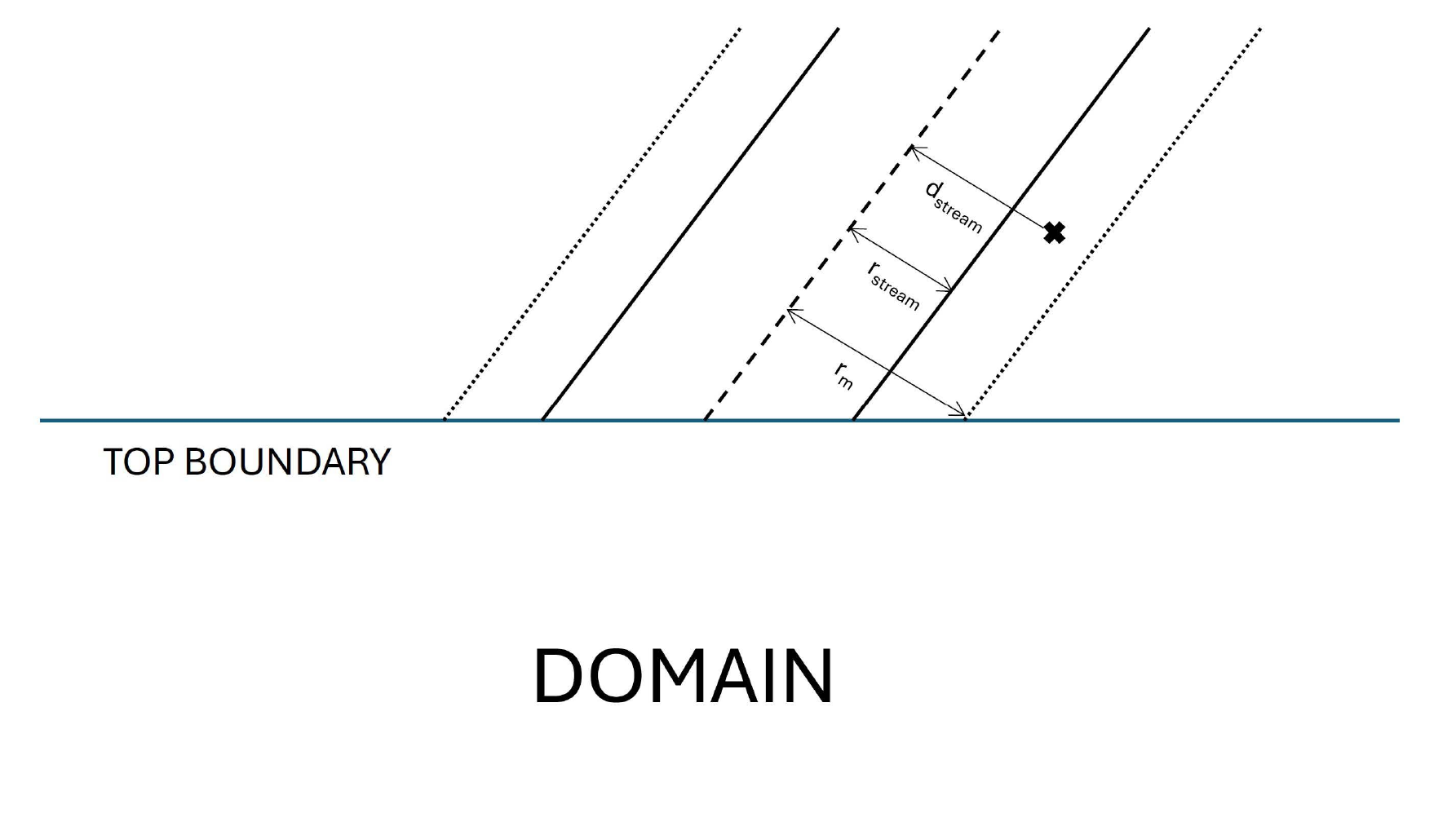}
\caption{Schematic showing how the mask is defined for the stream at the top boundary. }
\label{fig:plot_stream_bc}
\end{figure}

\begin{figure}[ht!]
\centering
\includegraphics[width=0.45\textwidth]{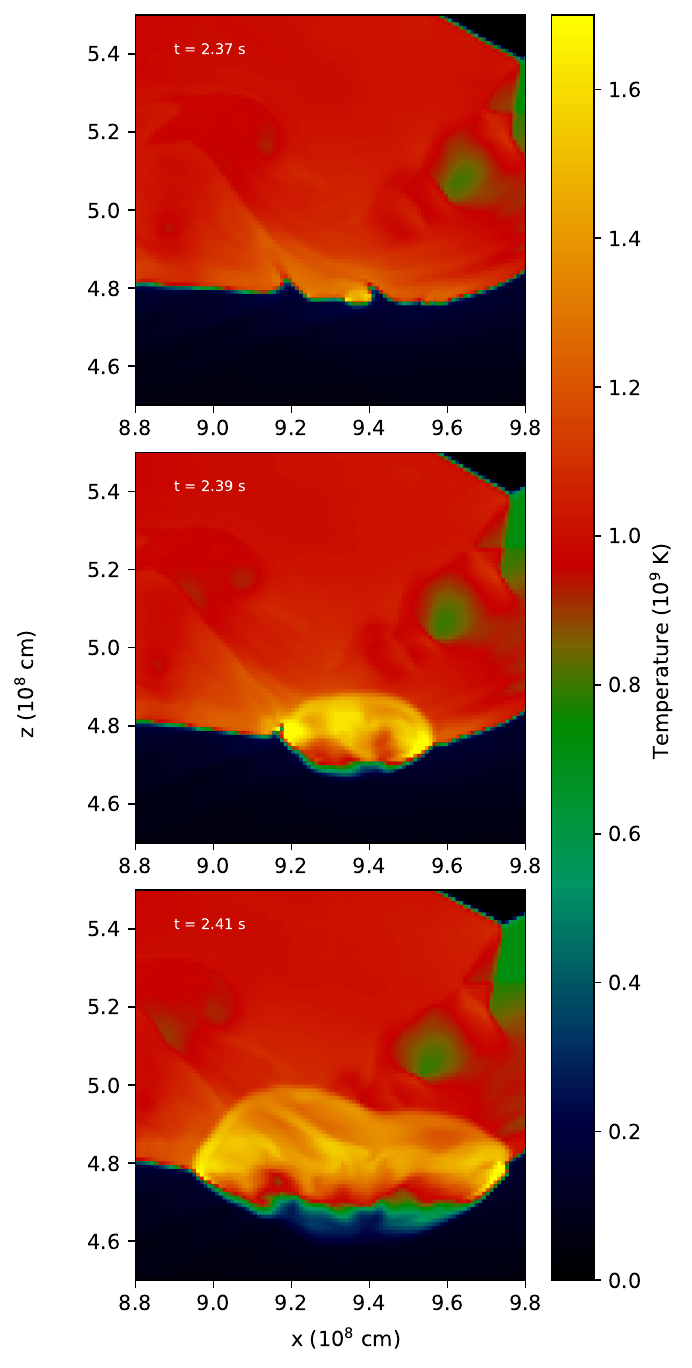}
\caption{Zoomed in snapshots of the ignition of a helium detonation in temperature for the case where a thick stream  impacts a 0.9 $\,M_\odot$ WD.}
\label{fig:plot9}
\end{figure}

\begin{figure*}[ht!]
\centering
\includegraphics[width=0.9\textwidth]{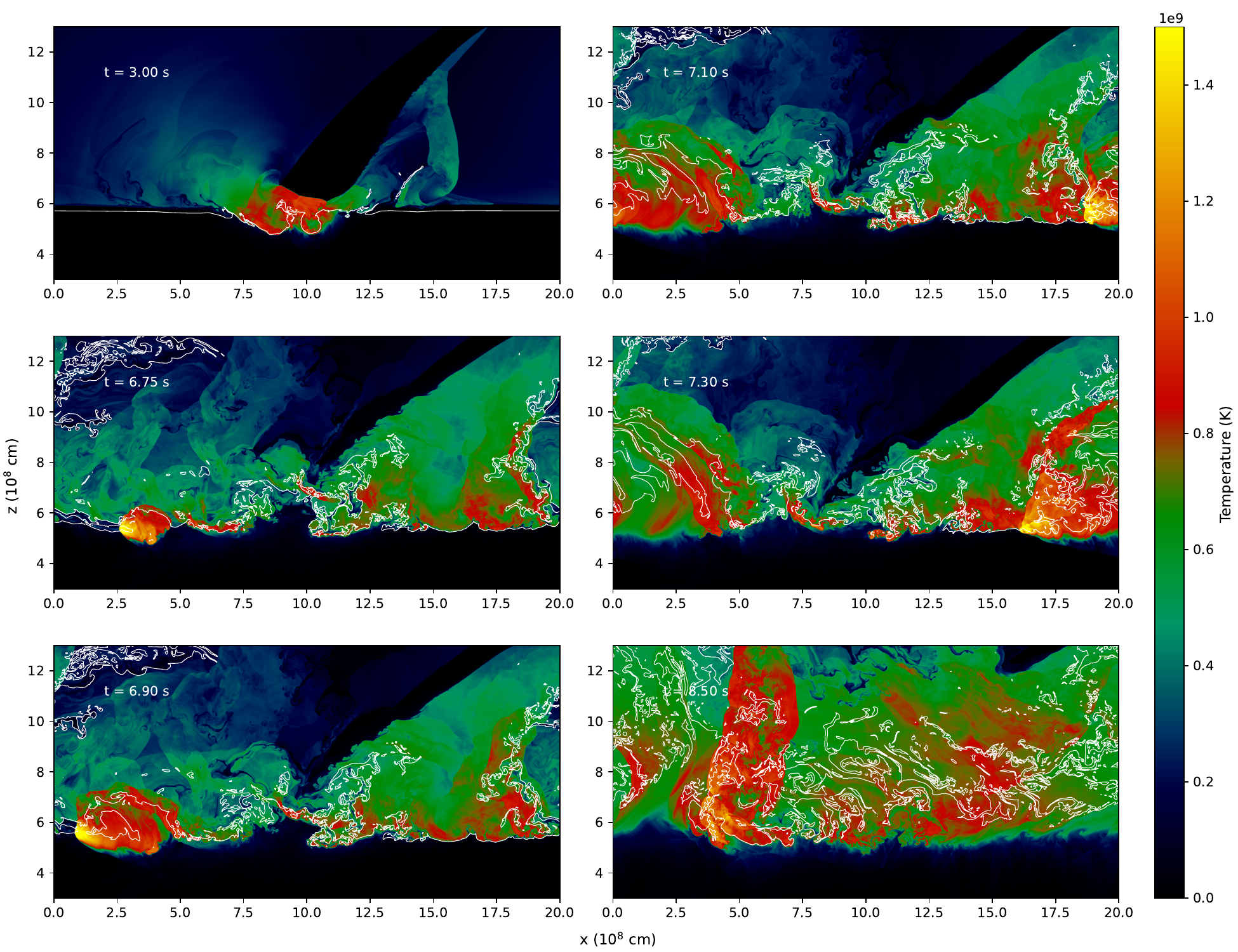}
\caption{Snapshots of temperature showing the ignition and propagation of a helium detonation for the case where a thin stream ($r_{\rm{stream}} = 0.5 \times 10^8$\ cm and $r_{\rm{m}} = 0.75 \times 10^8$\ cm) impacts a 0.9 $\,M_\odot$ WD. The ignition occurs away from the stream impact point and propagates to the left around the WD. The white contour lines connect points at which the $^{4}$He mass fraction equals the $^{12}$C mass fraction. In the top-left plot the white contour indicates the position of the initial transition region.}
\label{fig:plot1}
\end{figure*}

\begin{figure*}[ht!]
\centering
\includegraphics[width=0.9\textwidth]{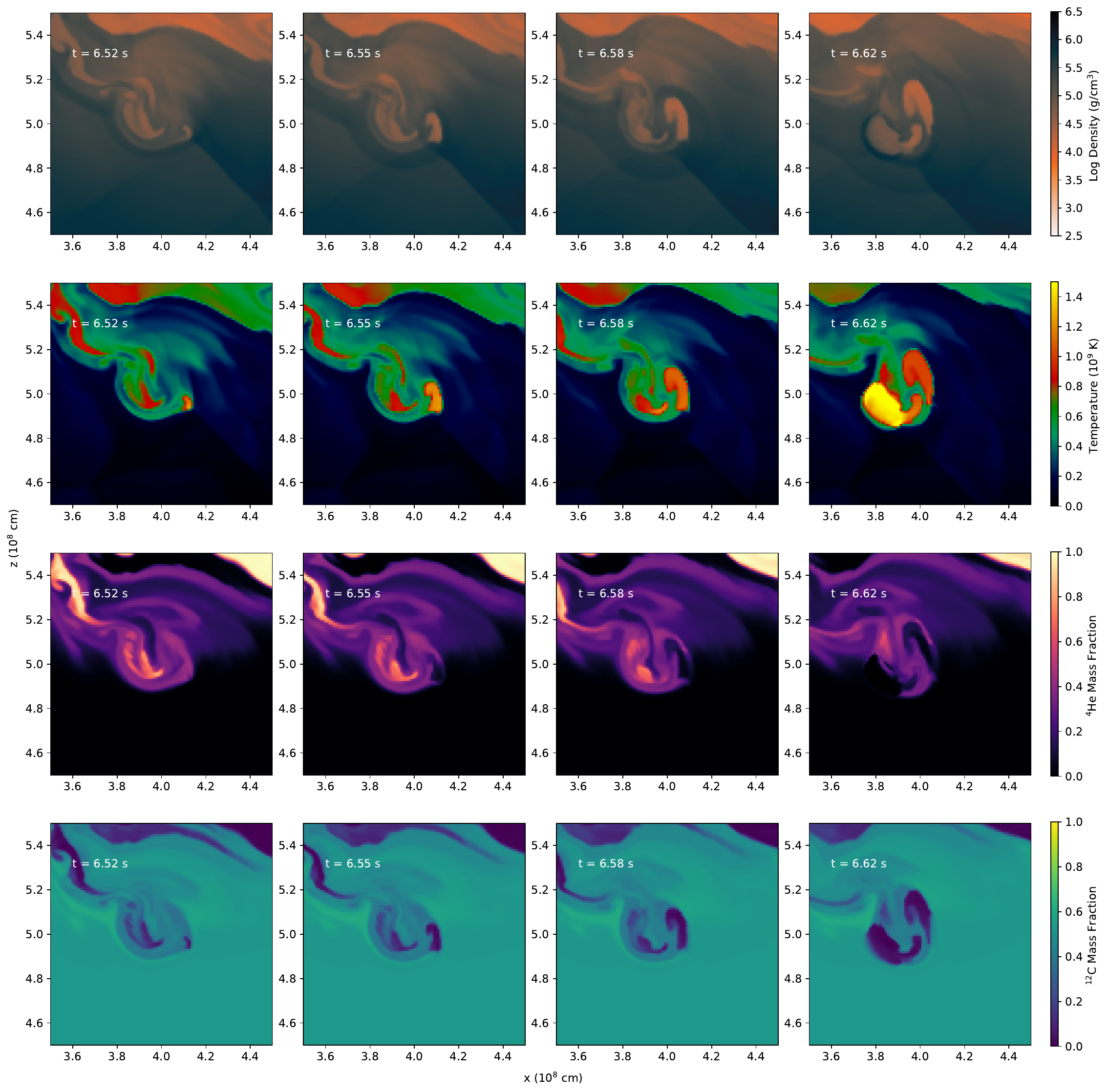}
\caption{Zoomed in snapshots of temperature, $^{4}$He mass fraction and $^{12}$C mass fraction at different times leading up to the ignition for the case where a thin stream impacts the surface of a 0.9 $\,M_\odot$ WD.}
\label{fig:plot8}
\end{figure*}

\begin{figure}[ht!]
\centering
\includegraphics[width=0.5\textwidth]{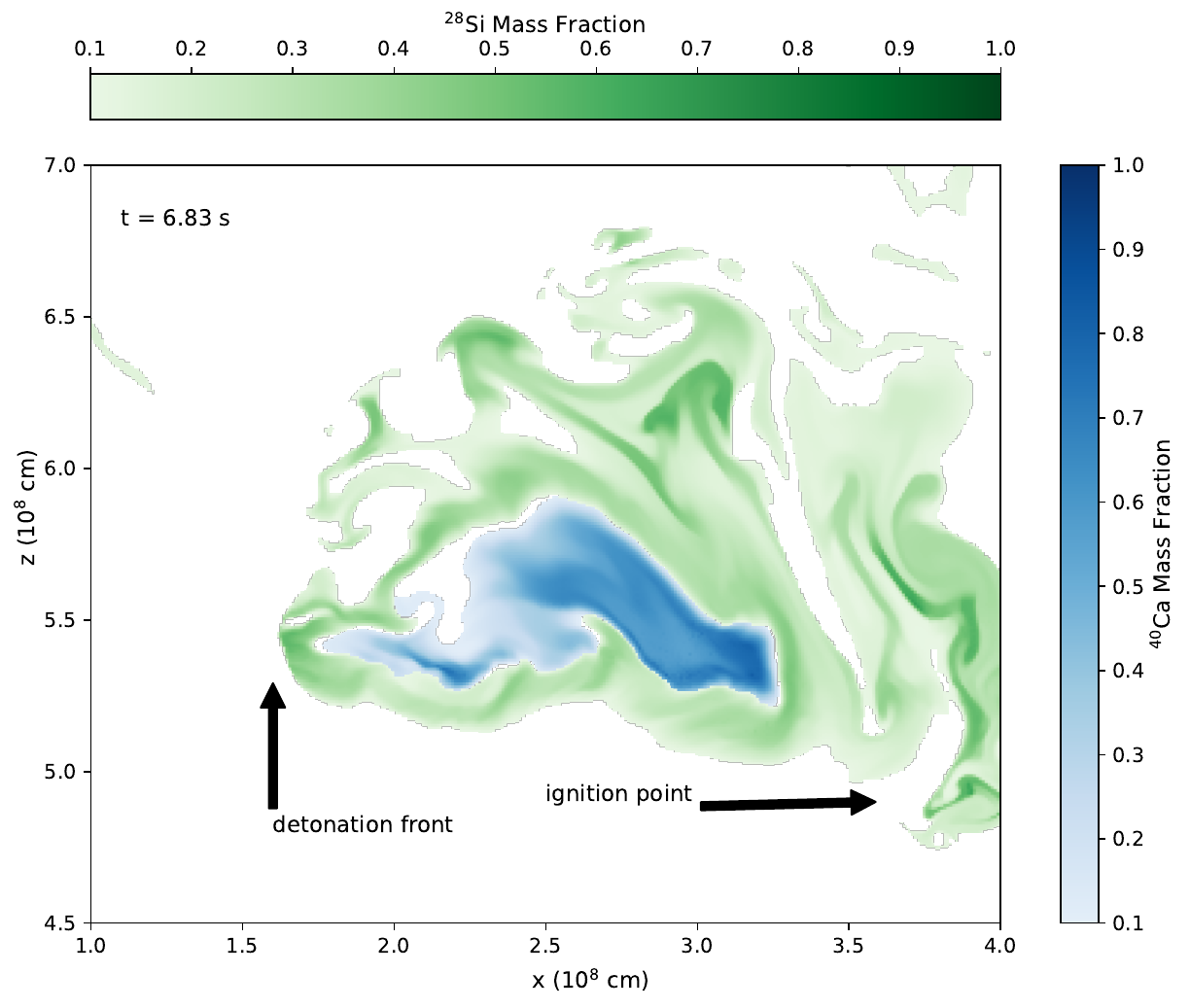}
\caption{$^{28}$Si and $^{40}$Ca mass fractions produced after the detonation propagates through a stratified region on the WD surface of a $^{4}$He-poor $^{12}$C-rich layer sandwiched by $^{4}$He-rich $^{12}$C-poor layers in the case where a thin stream impacts a  0.9 $\,M_\odot$ WD. The presence of such a stratified region causes the production of higher mass elements, mainly $^{40}$Ca, in the interior of the stratified region as the detonation passes through it and lower mass elements in the less dense regions. }
\label{fig:ca}
\end{figure}

\begin{figure*}[ht!]
\centering
\includegraphics[width=0.9\textwidth]{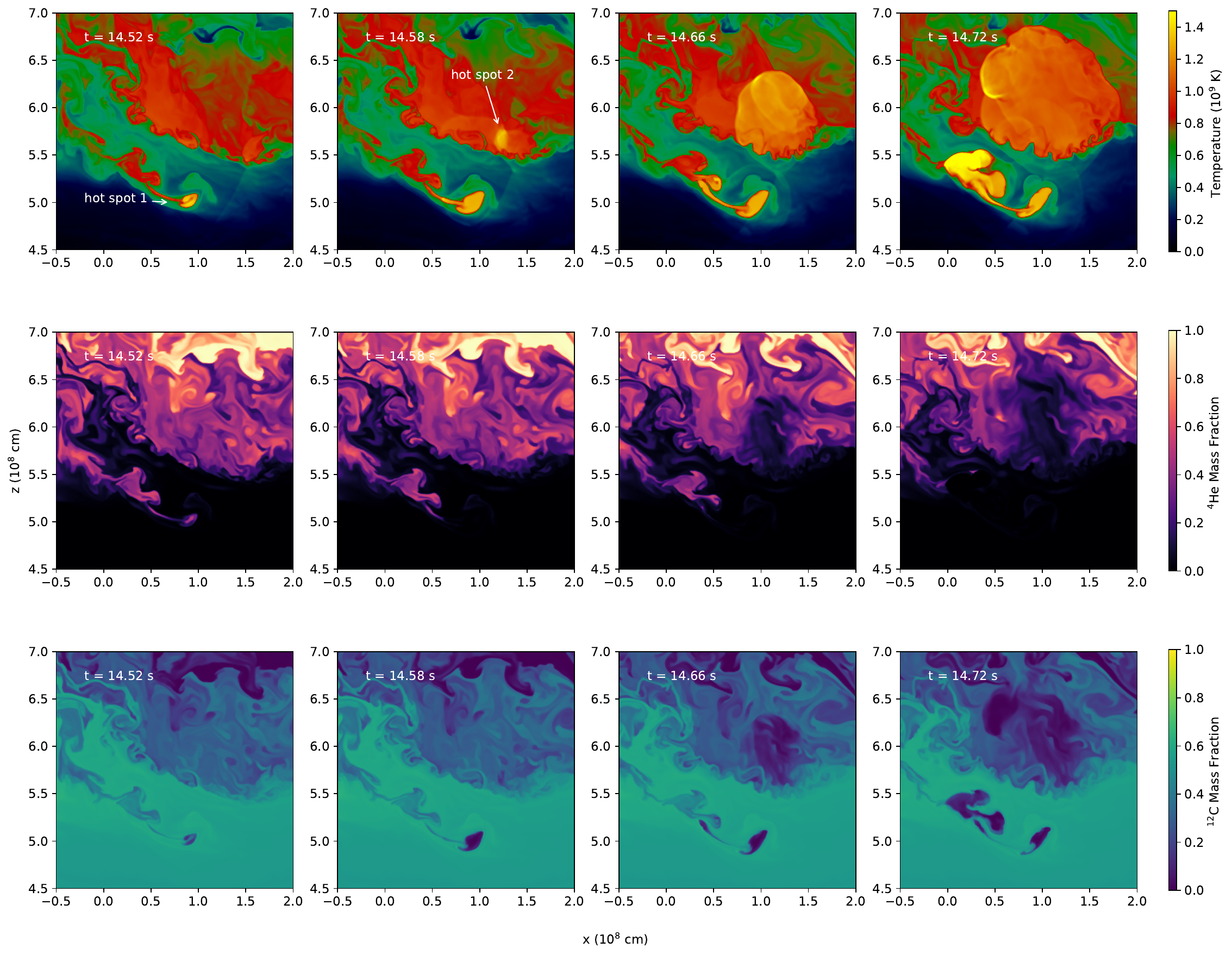}
\caption{Zoomed in snapshots of temperature, $^{4}$He mass fraction and $^{12}$C mass fraction at different times leading up to the ignition for the case where a thin stream impacts the surface of a 0.8 $\,M_\odot$ WD.}
\label{fig:plot7}
\end{figure*}

\begin{figure}[ht!]
\centering
\includegraphics[width=0.5\textwidth]{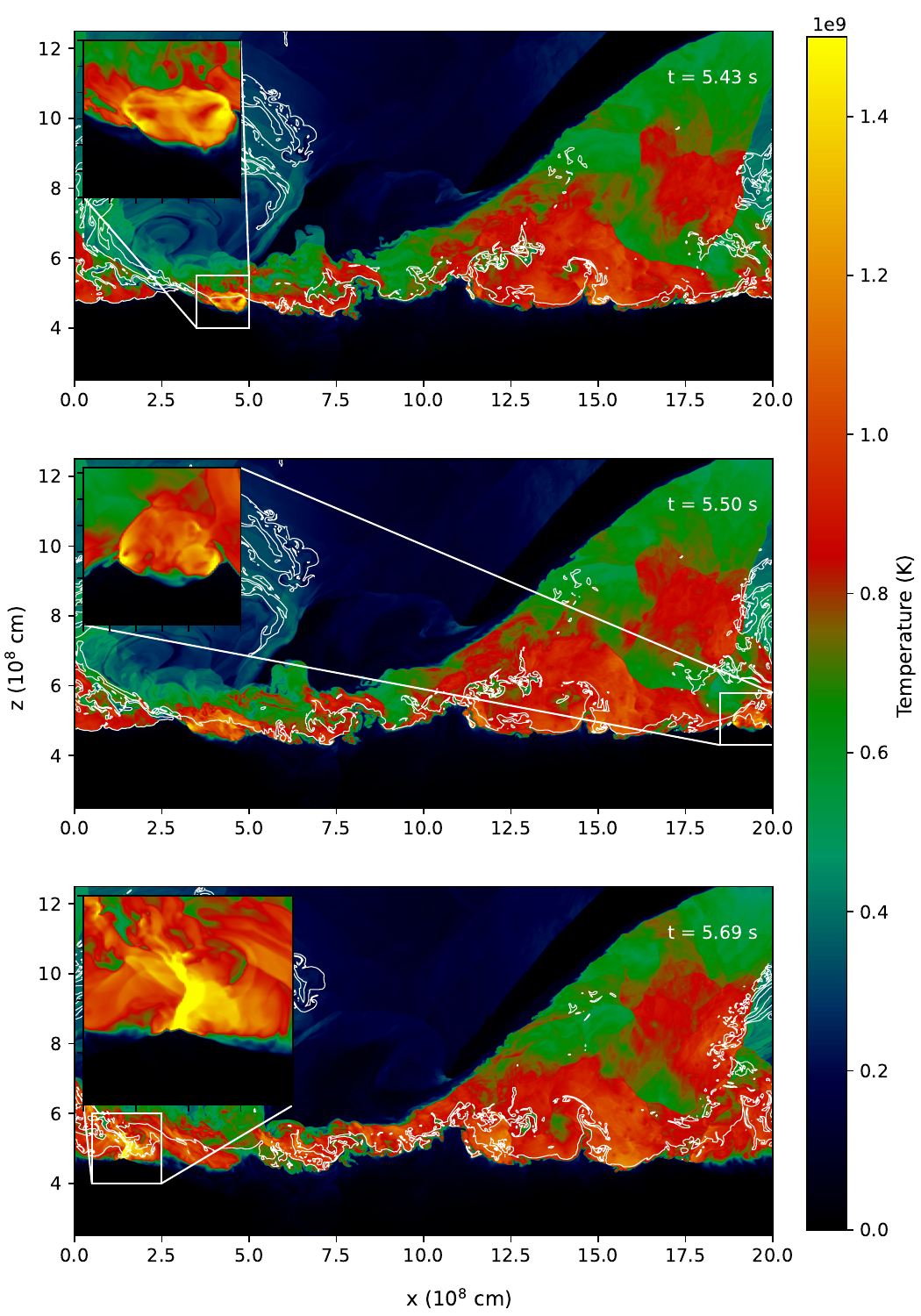}
\caption{Snapshots of temperature showing two ignitions produced one after the other in the case where a thick stream ($r_{\rm{stream}} =10^8$\ cm and $r_{\rm{m}} = 1.25 \times 10^8$\ cm) impacts a 1.0 $\,M_\odot$ WD. The helium detonations eventually collide and fizzle out. The white contour lines connect points at which the $^{4}$He mass fraction equals the $^{12}$C mass fraction.}
\label{fig:plot2}
\end{figure}

\begin{figure*}[ht!]
\centering
\includegraphics[width=0.9\textwidth]{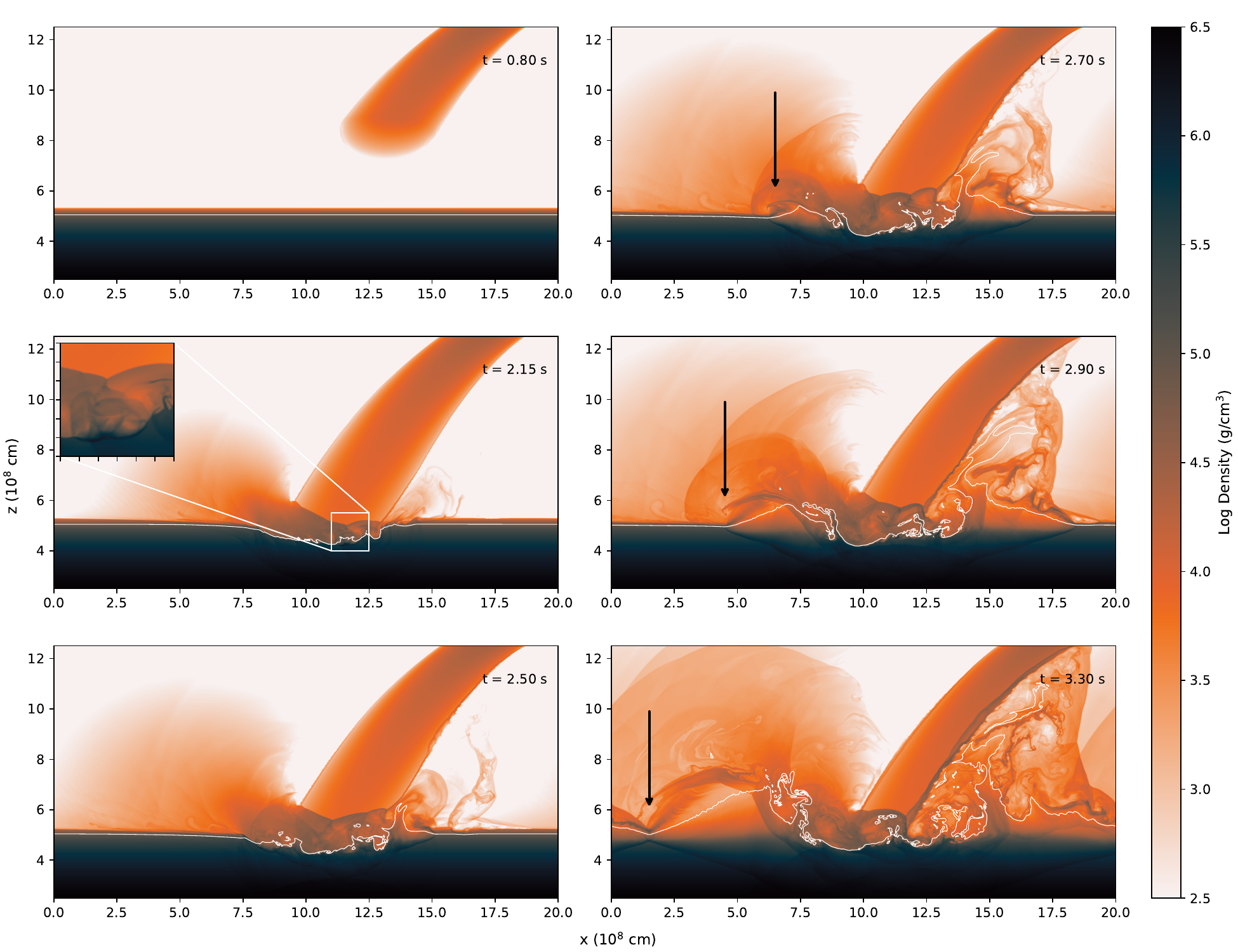}
\caption{Snapshots of log density for the 1.0 + 0.001 $\,M_\odot$ case where a thick stream impacts the surface. Ignition occurs at t$_{i}$ = 2.1 s and the resulting helium detonation propagates around the WD.The white contour lines connect points at which the $^{4}$He mass fraction equals the $^{12}$C mass fraction. The black arrows point to the propagating helium detonation.}
\label{fig:plot3}
\end{figure*}

\begin{figure}[ht!]
\centering
\includegraphics[width=0.45\textwidth]{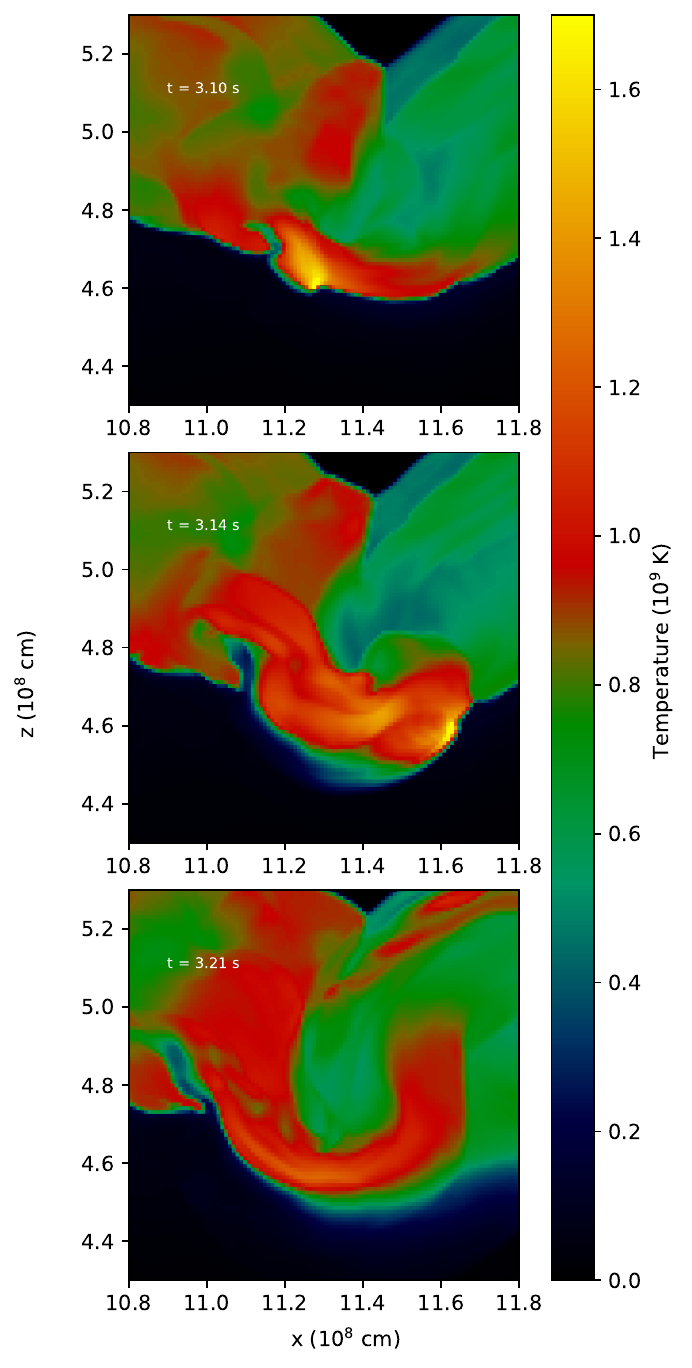}
\caption{Zoomed in temperature snapshots showing an example of a hot spot that ignites but immediately fails to propagate in the 1.0 + 0.001 $\,M_\odot$ case where a thin, lower density stream ($r_{\rm{stream}} = 0.5 \times 10^8$\ cm, $r_{\rm{m}} = 0.75 \times 10^8$\ cm and) impacts the surface of the WD. }
\label{fig:fail}
\end{figure}

\begin{figure}[ht!]
\centering
\includegraphics[width=0.45\textwidth]{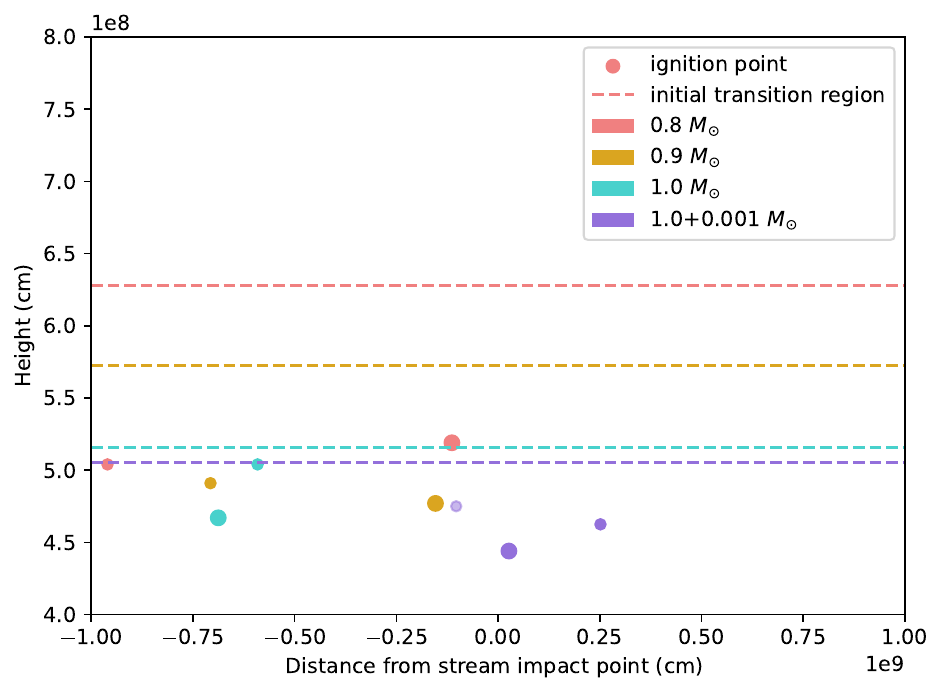}
\caption{Location of the ignition point in each of our simulations. The size of the point indicates the stream width, therefore large points are cases with the thick streams. The one case with a lower stream density is shown in a lighter color. This plot shows that the the ignition point for the helium detonation can lie anywhere on the surface and does not have to be near the stream impact point. }
\label{fig:plot4}
\end{figure}

\begin{figure}[ht!]
\centering
\includegraphics[width=0.5\textwidth]{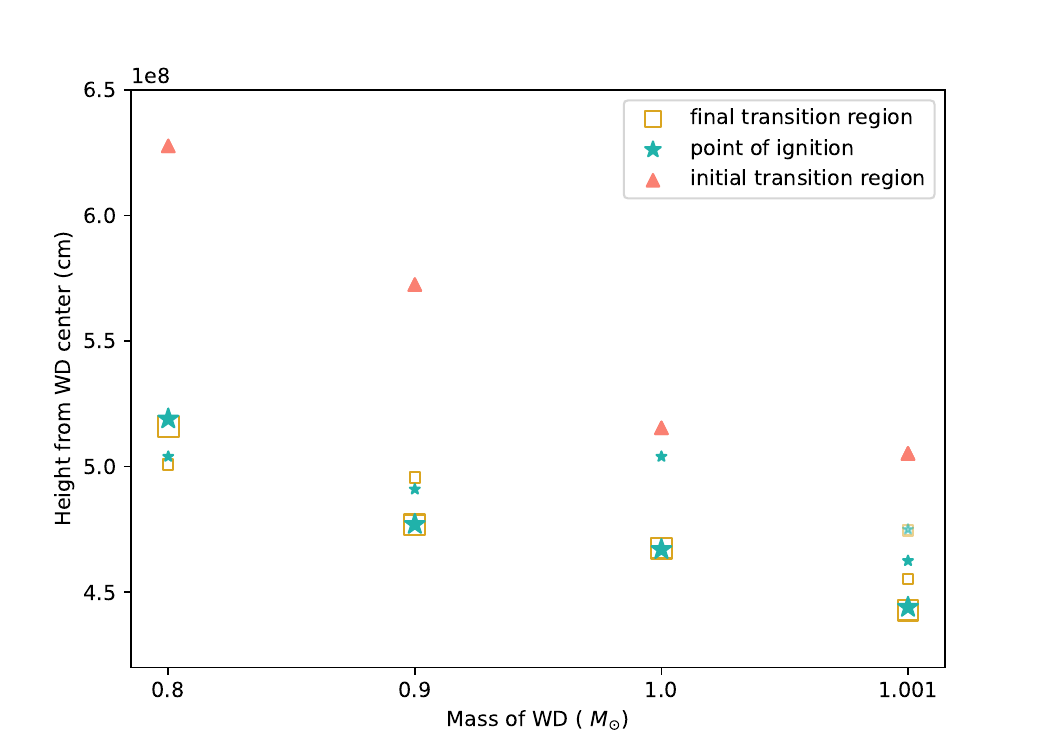}
\caption{Location of the ignition point with respect to the transition region. The transition region is defined as the vertical location at which the $^4$He and $^{12}$C abundances are equal.In the simulation, the stream impact and resulting addition of $^4$He mostly compresses the surface and moves the transition layer downward. The final position of the transition region just before ignition is indicated by open boxes, the initial position by triangles, and the height of the ignition by stars. The size of the point indicates the stream width, therefore large points are cases with the thick streams. The one case with a lower stream density is shown in a lighter color. This plot shows that the ignition happens very close to the transition region because of the presence of $^{12}$C mixed in with $^{4}$He and that ignition generally happens much deeper and at higher densities than the initial transition region location. Note that the final transition region for the 1.0 $\,M_\odot$ thin stream case is not included here since the location of the final transition region could not be determined accurately.}
\label{fig:plot5}
\end{figure}


\begin{figure}[ht!]
\centering
\includegraphics[width=0.5\textwidth]{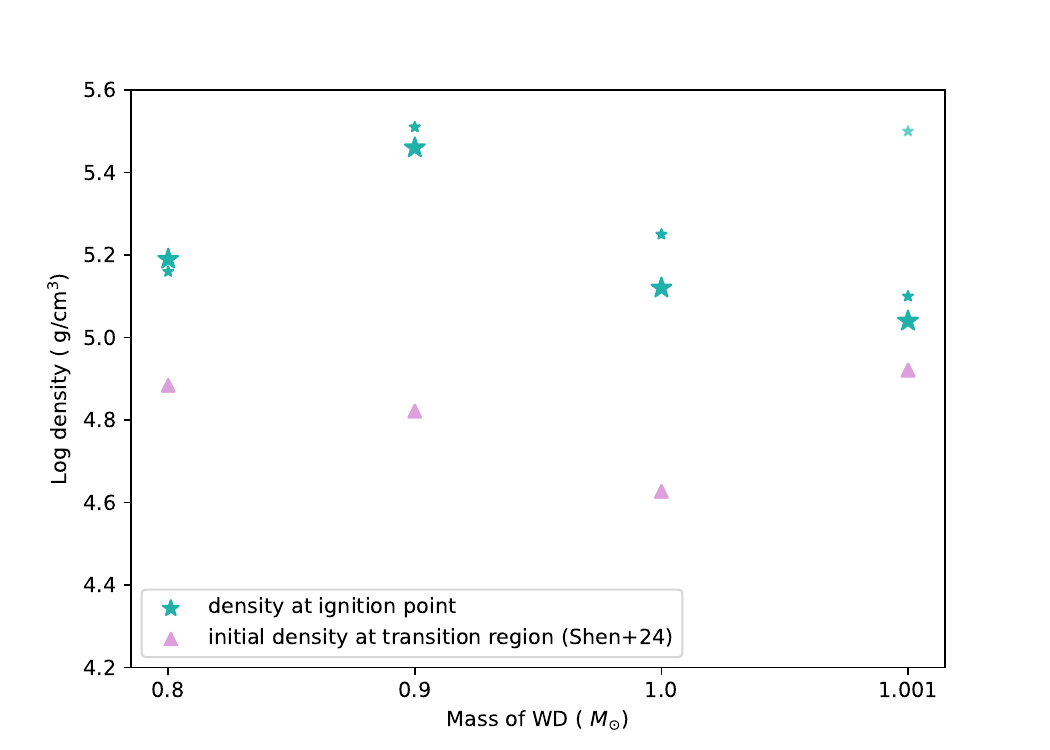}
\caption{Log density at the ignition point with respect to the ignition density from \citet{Shen_2024} for each of the WD masses. The size of the stars indicates the stream width, therefore large stars are cases with the thick streams. This plot shows that ignition occurred at higher densities for all cases compared to the undisturbed shells in \citet{Shen_2024}.}
\label{fig:plot6}
\end{figure}

\subsection{Modifications to the top boundary condition (stream)}

The stream from \citet{Rajavel2025} had a uniform density. This causes numerical issues in multi-dimensional runs at the stream edges where there is a sharp density, temperature and velocity gradient. The lateral density profile of the stream is now taken to be an inverted parabola, such that the maximum density occurs along the center of the stream, and the density goes to near zero at the stream edge, taken to be a distance
$r_{\rm{stream}}$ from the stream centerline. Outside that distance from the center, a density matching that of the fluff is used (0.1\ g\ cm$^{-3}$).  Both the material inside the stream and the low-density material outside it are given the same velocity along the stream direction. This low-density flowing material must then be blended with the usual boundary condition (typically zero-gradient). In order to accomplish that blending without creating numerical problems, we employed a two part solution: (1) we used a mask for the top boundary condition (from where the stream enters the domain) so that density, temperature and other stream parameters have a smooth transition from the stream to the fluff; and (2) we also defined a rectangular region around the stream to have the maximum resolution specified in our simulation. This again helps in preventing numerical issues.

At the beginning of each timestep, information from the domain is used to fill a layer of cells 4 layers deep just outside the computational domain. The hydrodynamic solver then uses that data to
compute the evolution (fluid flow) near the edge of the domain. Away from the stream we use a zero-gradient boundary condition, such that the first cell outside the domain uses the value of the last cell inside the domain. The mask then uses the zero-gradient boundary condition to set the value in each cell for each parameter. Figure \ref{fig:plot_stream_bc} shows a schematic of the applied mask for the top boundary condition. The mask is defined as so: 

\begin{enumerate}
    \item The shortest distance from a point on the boundary (guard) cells of the domain to the stream center, $d_{\rm{stream}}$, is calculated. 
    \item If $d_{\rm{stream}}$ is less than the half-width of the stream, $r_{\rm{stream}}$, then the mask value, m, is defined as 1.
    \item $r_{\rm{m}}$, the mask radius, is chosen as the the distance from the stream center greater than which mask value, m, is set to 0. Therefore, if $d_{\rm{stream}}$ is greater than $r_{\rm{m}}$ then the mask value is set to 0.    
    \item If $d_{\rm{stream}}$ is greater than $r_{\rm{stream}}$ but less than $r_{\rm{m}}$, the mask value is set to $|(r_{\rm{m}} - d_{\rm{stream}}) / (r_{\rm{m}} - r_{\rm{stream}})|$.
    \item We define each parameter using the following example equation:
    density = $\rho_{\rm{stream}} \times m$ + $\rho_{\rm{zero gradient}} \times (1-m) $
\end{enumerate}

Doing so ensures that we don't have any sharp gradients at the stream edges. 

\subsection{Choice of cases studied and parameters}

We simulate four different cases for WD mass: 0.8 $\,M_\odot$, 0.9 $\,M_\odot$, 1.0 $\,M_\odot$ and 1.0 + 0.001 $\,M_\odot$. The 1.0 + 0.001 $\,M_\odot$ is the MESA WD profile of a 1.0 $\,M_\odot$ WD with a 0.001 $\,M_\odot$ accreted layer. Each of these cases is run with two different stream half-widths which we will refer to as the ``thin" and the ``thick" stream from here on out. The thick stream has a stream half-width, $r_{\rm{stream}}$, of $10^8$\ cm and a $r_{\rm{m}}$ of $1.25 \times 10^8$\ cm. This is also the stream half-width used in \citet{Rajavel2025}. The thin stream has a $r_{\rm{stream}}$ of $0.5 \times 10^8$\ cm and a $r_{\rm{m}}$ of $0.75 \times 10^8$\ cm. The entry velocity of the stream was set to $6 \times 10^8$\ cm\ s$^{-1}$, entry angle to 35.54$^\circ$ and stream density to $2.5 \times 10^4$\ g\ cm$^{-3}$. The stream is composed of $^{4}$He. The stream parameters, like entry velocity and entry angle, were chosen based on our previous work \citep{Rajavel2025} where we integrated the trajectory of a test mass placed at L1 of a binary system when the radius of the companion WD is equal to the radius of the Roche lobe. The stream density and half-width were also chosen based on simulations from our previous work. See \citet{Rajavel2025} for more details on this. The domain width or the surface length is set to $2 \times 10^9$\ cm and not the circumference of the WD since increasing domain length increases the computation time but does not change if there is an ignition or not \citep{Rajavel2025}. The domain height is set to $10^9$\ cm. As a test to understand the effect that the stream density has on the helium ignition, we additionally simulated the 1.0 + 0.001 $\,M_\odot$ WD case with a thin ($0.5\times 10^8$\ cm half-width) stream and a lower $\rho_{\rm{stream}}$ of $1.25 \times 10^4$\ g\ cm$^{-3}$.

\section{Results} \label{sec:results}

According to \citet{Shen_2024}, all tested WD models with masses below 1.0 $\,M_\odot$ were able to sustain a propagating detonation when a hot spot was introduced at the core–shell transition region. For the 1.0 $\,M_\odot$ and 1.1 $\,M_\odot$ cases, an additional accreted layer (0.001 $\,M_\odot$ and 0.002 $\,M_\odot$, respectively) was required to propagate a helium detonation. They also found that models capable of sustaining a helium detonation also triggered a secondary core detonation.

A summary of the results from our study are documented in Table \ref{tab:deluxesplit}. Our results show that all simulated cases ignite a helium detonation though not all of these detonations propagated successfully. In some models, ignition occurs at the stream impact point, whereas in others it occurs significantly away from the impact location. Since each of our simulated cases differ from each other, they will be discussed separately in the subsections below but will be discussed out of order for better continuity. An overview and a comparison of results will be discussed in section \ref{sec:discussion}. Animated plots of all cases are available via the Zenodo link provided at the end of the manuscript.





\subsection{0.9 $\,M_\odot$  WD}

The thick stream ignites a helium detonation on the surface of the 0.9 $\,M_\odot$ WD at $t = 2.37$ s, manifesting the ``direct mechanism'' observed in \citet{Rajavel2025}, where ignition occurs at the stream impact point. After ignition, the detonation propagates only to the left of the ignition point, while propagation to the right quickly stalls. The peak density at the ignition point is listed in Table \ref{tab:deluxesplit}.

Figure \ref{fig:plot9} shows the helium ignition in temperature. The initial stream impact compresses the helium shell deeper into the surface, forming a pit and increasing the density at the transition region where ignition ultimately occurs. Since the detonation requires helium for $\alpha$-capture to propagate,  it appears to climb the walls of this pit before propagating along the transition region between the helium shell and the carbon–oxygen core. 

The main products formed behind the detonation are $^{28}$Si and $^{32}$S, which are produced with the highest abundances. Small amounts of $^{24}$Mg and $^{20}$Ne are also synthesized behind the detonation (mass fractions $> 0.1$), along with even smaller amounts of $^{36}$Ar, $^{40}$Ca (mass fractions $> 0.1$ in some parts). Small amounts of $^{44}$Ti (mass fraction between 0.01 and 0.1) is also produced, but no elements heavier than $^{44}$Ti are formed in significant quantities (mass fraction $> 0.01$).

For the thin stream case, a ``pop'' of helium burning occurs on the surface of the 0.9 $\,M_\odot$ WD at $t = 2.7$ s, but this does not develop into a propagating detonation. This burning near the stream impact point fizzles roughly 0.3 s after it occurs. At $t = 6.61$ s, ignition occurs again at a point to the left of the stream impact point. Snapshots of temperature in Figure \ref{fig:plot1} depict the ignition and propagation of the helium detonation. Figure \ref{fig:plot8} zooms into the ignition point in log density, temperature, $^{4}$He mass fraction and $^{12}$C mass fraction. The hot spot necessary for the ignition seems to occur in an isolated, high temperature, $^{4}$He-rich pocket. The expansion of this hot spot caused by $^{4}$He burning, compresses a second, neighboring $^{4}$He-rich, high temperature pocket next to it which leads to the formation of a second hot spot. $^{4}$He burning due to the second hot spot travels along the tendril of $^{4}$He. 

Ignition in this case involves two novel features. First, the transition of a compression wave into a detonation occurs within the composition transition region, with the process coming about due to the profile of $^{12}$C enrichment within that region. Secondly, ignition takes place as a two step process in which the compression wave that transitions into the detonation is generated by a previous burst of burning that is nearby but not contiguous in composition with the region in which the detonation is fully formed. The 0.9 $\,M_\odot$ WD profile exhibits a peak in $^{12}$C abundance at the outer edge of the core, just before the transition to the $^{4}$He-rich shell. As discussed more broadly across all cases in Section \ref{sec:discussion}, the impact of the accretion stream compresses, and in some places folds, the surface layers, pushing material deeper into the WD. As a result, the transition region, along with the $^{12}$C abundance peak, is shifted inward.

In Figure \ref{fig:plot8}, two $^{4}$He-rich pockets appear to form near the transition layer. It appears that when the small portion of the core-shell boundary was pushed deeper into the core material, the slightly $^{12}$C-rich region was carried inward with it. This $^{12}$C-rich region can be seen in the first panel in the bottom row of Figure \ref{fig:plot8}, outlining the $^{4}$He-rich region on the bottom side. Compression waves passing through the pocket on the right create a hot spot that causes burning and expansion of material. The compression wave from the expansion of this pocket causes another hot spot to be created in the neighboring pocket on the left. The actual ignition occurs when the high temperature passes over the $^{12}$C-rich region, after which the resulting detonation propagates along the $^{12}$C-rich layer. We view this sequence as representative of the complex process that can lead to a detonation in a realistic layer.

As the detonation propagates, it crosses a region with a stratified composition in $^{4}$He, $^{12}$C, and $^{16}$O. Immediately above the carbon abundance peak lies a layer with relatively low $^{12}$C and $^{16}$O abundances but higher $^{4}$He content. Above this is a layer with higher $^{12}$C and $^{16}$O abundances and lower $^{4}$He, followed by an outer region dominated by helium. This stratified structure can be seen in Figure \ref{fig:plot8} in the leftmost $^{4}$He and $^{12}$C plots. The stratified structure likely arises from the dredge-up of $^{12}$C and $^{16}$O due to the stream impact and subsequent movement of the dredged up material across the WD surface.

As the detonation propagates through this stratified region, carbon and oxygen are efficiently burned, while the helium-rich, carbon-poor layer retains some unburned helium. Lighter $\alpha$-chain elements, such as $^{20}$Ne, $^{24}$Mg, and $^{28}$Si, are synthesized in the outermost regions of the ash behind the detonation front, whereas $^{32}$S and $^{36}$Ar form in intermediate layers of the ash. Heavier elements, including $^{40}$Ca, $^{44}$Ti, and $^{48}$Cr, are produced in the interior regions of the ash.

Figure \ref{fig:ca} highlights the regions where $^{40}$Ca and $^{28}$Si are synthesized after the detonation has propagated to the left. The interior region of the ash refers to the region where $^{40}$Ca is predominantly produced as seen in Figure \ref{fig:ca}. The outermost region of the ash refers to the region where $^{28}$Si is mainly produced and the intermediate region (not shown in Figure \ref{fig:ca}) is the region between the interior and outermost region of the ash where $^{32}$S and $^{36}$Ar are produced. Notably, the interior region that predominantly produces $^{40}$Ca is also the region that retains unburned helium and lacks lighter $\alpha$-chain elements. This suggests that $\alpha$-capture onto $^{12}$C, $^{16}$O, and other lighter $\alpha$-chain nuclei is the dominant burning process in this region. In particular, $^{40}$Ca is produced almost exclusively in this layer, reaching mass fractions as high as $\sim 0.8$, and is not significantly synthesized elsewhere along the surface.

The other primary products behind the detonation front are $^{28}$Si, $^{32}$S, and $^{36}$Ar. Additionally, $^{20}$Ne and $^{24}$Mg are produced with mass fractions $> 0.1$. Very small amounts of $^{44}$Ti and $^{48}$Cr are also produced, with mass fractions between 0.1 and 0.001. No elements heavier than $^{48}$Cr are produced in significant quantities.

Ultimately, the detonation propagates to the left of the domain while failing to propagate on the right. This detonation is able to traverse the entire WD surface once. Following the first pass, a shockwave continues to propagate around the WD for a second time, but fizzles out roughly halfway around the star. 

To test the sensitivity of our results with respect to resolution, we ran a lower resolution simulation (cell size $\approx 20$ km) of the thin stream case. A helium detonation ignited at $t = 5.71$ s but does not follow the same mechanism as the $\approx 10$ km resolution simulation. The ignition mechanism is similar to that seen in Figure \ref{fig:plot9}. The ignition location also varies from the higher resolution simulation (see Table \ref{tab:deluxesplit}). We also restarted the $\approx 10$ km resolution simulation just before ignition at a lower resolution of $\approx 20$ km to see if the ignition follows the same mechanism at a lower resolution in the presence of the two pocket structure. We see that at a lower resolution the right pocket seen in Figure \ref{fig:plot8} ignites a helium detonation. As expected, these low resolution simulations show that different amounts of resolution dependent mixing can cause the formation of distinct structures. Due to this, as the resolution is varied, the surface may ignite earlier or later through different mechanisms. This also shows that a higher resolution simulation than the ones we have performed may ignite at a different location and through a different mechanism and at a different time.

\subsection{0.8 $\,M_\odot$  WD}

We performed two simulations of the 0.8 $\,M_\odot$ WD that differ only in stream thickness. As mentioned previously, both cases ignited and developed a propagating detonation. 

In the thick stream case, ignition occurs shortly after the stream impacts the surface, at $t = 2.48$ s. The detonation propagates to both the left and right of the ignition point and eventually wraps around the surface of the star. This mechanism is similar to the direct mechanism. 

Examination of the mass fractions of the elements in the nuclear reaction network during the propagation of the detonation shows depletion of $^4$He and $^{12}$C. $^{12}$C is mostly consumed in the He-rich region while leaving some $^4$He. Regions in which the two have similar abundances, either in the interface layer or mixed regions, appear to consume $^{4}$He, leaving some $^{12}$C. As the detonation propagates, the primary products just behind the detonation are $^{28}$Si and $^{32}$S. Additionally, smaller amounts of $^{36}$Ar, $^{40}$Ca, $^{20}$Ne, and $^{24}$Mg are also produced just behind the detonation with mass fractions $> 0.1$. Small amounts of $^{44}$Ti and $^{48}$Cr are also produced behind the detonation with mass fractions between 0.01 and 0.1 in the first 0.5 s after ignition. No elements heavier than $^{48}$Cr are synthesized in significant quantities (mass fraction $> 0.01$).

Additionally, the burned material exhibits a stratified composition. The outermost regions of the ash are dominated by lighter $\alpha$-chain elements such as $^{20}$Ne, $^{24}$Mg, $^{28}$Si and $^{32}$S. $^{28}$Si and $^{32}$S also dominate the intermediate region while the interior region of the ash contains heavier $\alpha$-chain products similar to Figure \ref{fig:ca}. This interior region of the ash is also completely depleted of $^{12}$C. Such a stratified structure suggests that $\alpha$-capture onto lighter $\alpha$-chain elements plays a key role in producing heavier elements like $^{36}$Ar and $^{40}$Ca.

In the thin stream case, ignition occurs at $t = 14.5$ s. The stream continuously deposits material onto the WD surface, which is deflected after impacting the WD surface and continues to wrap around the surface of the star. This lateral flow of material eventually interacts with the incoming stream and also disrupts and compresses the existing helium shell. By $t \approx 14$ s, the helium shell is highly disturbed. Although, several localized temperature spikes exceeding $10^{9}$ K occur during the simulation, these events produce only brief, localized burning and do not develop into detonations. 

At $t = 14.50$ s, an ignition occurs far left of the stream impact point within the disrupted surface layer. On close examination, there seem to be two hot spots formed one after the other as seen in Figure \ref{fig:plot7} at t = 14.50 and t = 14.57 s respectively. The first hot spot (labeled 1 in Figure \ref{fig:plot7}) forms within a $^{4}$He-rich, high temperature pocket surrounded by $^{12}$C-rich WD material. The flow of the accreted material on the surface causes turbulence that mixes material from the WD shell deeper into the WD surface leading to the formation of such a pocket. Hot spot 1 appears to be triggered by the impact of a compression wave also originating from the flow of mass on the disturbed surface. Hot spot 1 is hot enough for helium burning to occur and the burning travels along the $^{4}$He-rich tendril connected to this pocket where it eventually ignites into a helium detonation.

Around the same time, a second hot spot (labeled 2 in Figure \ref{fig:plot7}) is formed and also ignites into a propagating detonation though this detonation does not seem as strong as detonation 1 (formed from hot spot 1) due to the low temperature peak at the detonation front and the presence of unburnt helium in its wake. These propagating detonations eventually merge into one strong helium detonation. It is unclear if detonation 2 would have been able to propagate around the WD if detonation 1 did not occur. 

The resulting helium detonation propagates to the left of the ignition point through the disturbed surface and wraps around the entire WD surface. This behavior indicates that helium ignition can occur even after the surface has been substantially disrupted. The main products behind the detonation are $^{28}$Si and $^{24}$Mg. Small amounts of $^{20}$Ne, $^{32}$S and $^{36}$Ar, and even smaller amounts of $^{40}$Ca are also produced behind the detonation with mass fractions of $> 0.1$. No elements heavier than $^{40}$Ca are produced in significant quantities (mass fraction $> 0.01$). Note that the abundances are not all produced in the same region, and the mass fractions stated here describe variations between adjacent regions in a non-uniform composition of ash.

\subsection{1.0 $\,M_\odot$  WD}

The 1.0 $\,M_\odot$ case was one of the WD masses that failed to sustain a surface detonation in \citet{Shen_2024}. A hot spot introduced in the transition region was unable to propagate a detonation even at a minimum resolution of 3.17 km in their work. Our simulations show that the thin and thick stream ignite a detonation in the 1.0 $\,M_\odot$ WD; however, the detonation does not propagate completely around the WD surface.

In the thick stream case, two separate points on the WD surface ignite almost simultaneously. The first ignition occurs at $t = 5.36$ s while the second occurs at $t = 5.44$ s as shown in Figure \ref{fig:plot2}. The first detonation propagates only to the left, while the second propagates only to the right. Both ignitions occur much deeper than the initial transition region (where the carbon and helium abundances are equal), in a denser and higher-pressure region of the WD. This is likely because the accreted material from the stream compresses the transition region, increasing the local density and pressure. Consequently, the scale height at the ignition points ($\gtrsim 10^{7.5}$ cm) is larger than the scale height calculated in \citet{Shen_2024} ($\sim 10^{7}$ cm), enabling ignition. Despite ignition at two points, these detonations collide and fizzle immediately after.

In the thin stream case, ignition occurs at $t = 4.97$ s to the left of the stream impact point. The resulting detonation propagates leftward and propagates a little more than halfway around the WD surface before fizzling out near to the stream impact point. By this stage, the surface has been significantly disrupted, and the detonation propagates through a highly turbulent medium. As a result, the overall propagation is less coherent compared to cases with earlier ignition times ($t_i \leq 3$ s), where the surface remains relatively undisturbed. 

In this case too, the detonation propagates in the $^{12}$C abundance peak inherent to the WD profile. Above this peak lies a helium-rich, $^{12}$C and  $^{16}$O-poor layer, followed by a $^{12}$C and  $^{16}$O-rich layer formed through dredge-up and subsequent mixing driven by the stream impact. As the detonation propagates through these stratified layers, carbon and oxygen are completely consumed, while some helium remains unburned in the $^{12}$C and  $^{16}$O-poor region. This indicates that $\alpha$-capture on to $^{12}$C and $^{16}$O is the dominant process responsible for producing most of the $^{28}$Si, $^{32}$S and $^{36}$Ar.

The main products formed behind the detonation are $^{28}$Si and $^{32}$S. Additionally, $^{20}$Ne and $^{24}$Mg are produced with mass fractions $> 0.1$. Smaller amounts of $^{36}$Ar are also synthesized (mass fraction $> 0.01$), while no heavier elements are produced in significant quantities.

\subsection{1.0 $\,M_\odot$  WD with a 0.001 $\,M_\odot$  accreted layer}

Since the 1.0 $\,M_\odot$ WD was unable to propagate a detonation around the WD in \citet{Shen_2024}, the authors constructed a 1.0 $\,M_\odot$ WD with a 0.001 $\,M_\odot$ accreted layer in MESA and used the resulting WD profile to test whether detonation propagation was possible, unlike the plain 1.0 $\,M_\odot$ case. They found that the 1.0 $\,M_\odot$ WD with a 0.001 $\,M_\odot$ (henceforth, 1.0 + 0.001 $\,M_\odot$) accreted layer was able to sustain a helium detonation on its surface. Therefore, we also tested whether a stream could ignite a propagating helium detonation in this 1.0 + 0.001 $\,M_\odot$ configuration by simulating thick, thin, and thin lower-density streams. Similar to \citet{Shen_2024}, all of our 1.0 + 0.001 $\,M_\odot$ runs successfully ignited and propagated a helium detonation around the WD surface.

In the thick stream case, ignition occurs at $t = 2.1$ s via the direct mechanism, where the stream triggers a detonation shortly after impacting the WD surface. Log density snapshots at different points in time depicting the ignition and detonation propagation are shown in Figure \ref{fig:plot3}. The resulting helium detonation propagates in both directions, wrapping around the WD surface and colliding on the opposite side. A comparison of Figures \ref{fig:plot1} and \ref{fig:plot3} reveals the stark difference in how the detonations propagate on an undisturbed vs a disturbed surface.

The primary products formed behind the detonation front are $^{28}$Si and $^{32}$S, with small amounts of $^{20}$Ne and $^{24}$Mg (mass fractions $> 0.1$). Smaller amounts of $^{36}$Ar and $^{40}$Ca are also produced (mass fraction $> 0.01$). No elements heavier than $^{40}$Ca are synthesized in significant quantities. 

In the thin stream case, ignition occurs at $t = 4.73$ s. Unlike previous cases, which ignited either at or to the left of the stream impact point, this ignition occurs to the right of the stream impact point. The helium detonation propagates predominantly to the right, wrapping around the entire WD surface.  During propagation, a region forms in which helium is not completely consumed. Instead, $^{12}$C, $^{16}$O, $^{20}$Ne, and $^{24}$Mg are depleted, producing heavier elements such as $^{40}$Ca, along with small amounts of $^{44}$Ti and $^{48}$Cr (mass fractions between $ 0.01$ and 0.1). The dominant products remain $^{32}$S and $^{28}$Si, with smaller quantities of $^{36}$Ar, $^{24}$Mg, and $^{20}$Ne (mass fractions greater than 0.1). 

Finally, we performed an additional simulation with the thin stream while reducing the stream density by half to test whether detonation could still be triggered at a lower mass transfer rate. In this case, ignition occurs at $t = 3.4$ s by the direct mechanism and the detonation propagates to the left around the WD surface. As the detonation propagates, we observe the formation of two different layers. The main detonation front efficiently burns helium into $^{20}$Ne, $^{24}$Mg and $^{28}$Si, while a secondary layer near the leading edge exhibits complete depletion of $^{12}$C, $^{16}$O, $^{20}$Ne, and $^{24}$Mg, producing $^{32}$S, $^{28}$Si, $^{36}$Ar, and $^{40}$Ca, along with small amounts of $^{44}$Ti (mass fraction between 0.1 and 0.01). Overall, the dominant products are again $^{32}$S and $^{28}$Si, with smaller amounts of $^{20}$Ne, $^{24}$Mg, $^{36}$Ar, and $^{40}$Ca (mass fraction greater than 0.1), although the total yields of $^{36}$Ar, and $^{40}$Ca are comparatively lower. No elements heavier than $^{44}$Ti are produced in significant amounts (mass fraction greater than 0.01). At $t = 3.1$ s, a hot spot appears to form as shown in Figure \ref{fig:fail}. This ignition also seems to propagate a detonation until $t = 3.2$ s, after which we see that the detonation fizzles out and fails to propagate. This shows that even though an ignition can occur, the right conditions are required for the detonation to continue to propagate. 

\begin{deluxetable}{lccccccccc}
\tabletypesize{\scriptsize}
\tablewidth{0pt} 
\tablecaption{Summary of cases explored.\label{tab:deluxesplit}}
\tablehead{
    \colhead{WD Mass} & \colhead{$r_{\rm{stream}}$}& \colhead{$\rho_{\rm{stream}}$} & \colhead{$t_{i}$} &
    \colhead{$\rho_{i}$} & \colhead{$d_{i}$} & \colhead{$H_{i}$} & \colhead{$d_{tr1}$} & \colhead{$d_{tr2}$} & \colhead{Propagation Direction}\\
    \colhead{($\,M_\odot$)} & \colhead{($10^8$\ cm)} & \colhead{($10^4$\ g\ cm$^{-3}$)} & \colhead{(s)} & \colhead{($10^5$\ g\ cm$^{-3}$)} & \colhead{($10^8$\ cm)}& \colhead{($10^8$\ cm)}& \colhead{($10^8$\ cm)} & \colhead{($10^8$\ cm)} & \colhead{} \\
} 
\startdata 
0.8 & 1 & 2.5 & 2.48 & 1.55 & -1.14 & 5.19 & 6.28 & 5.16 & both\\ 
0.8 & 0.5 & 2.5 & 14.5 & 1.45 & -9.60 & 5.04 & 6.28 & 5.01 & left\\ \hline
0.9 & 1 & 2.5 & 2.37 & 2.88 & -1.54 & 4.77 & 5.73 & 4.77 & left\\ 
0.9 & 0.5 & 2.5 & 6.6 & 3.24 & -7.07 & 4.91 & 5.73 & 4.96 & left\\ 
0.9\** & 0.5 & 2.5 & 5.71 & 3.29 & -2.62 & 5.36 & 5.73 & N/A & left\\ \hline
1.0 & 1 & 2.5 &  5.36/5.44 & 1.32 & -6.88/7.79 & 4.67/4.97 & 5.15 & 4.68 & \***\\ 
1.0 & 0.5 & 2.5 & 4.97  & 1.79 & -5.91 & 5.04 & 5.15 & N/A & left\\ \hline
1.0 + 0.001 & 1 & 2.5 & 2.1 & 1.10 & 0.27 & 4.44 & 5.05 & 4.43 & both\\ 
1.0 + 0.001 & 0.5 & 2.5 &  4.73 & 1.26 & 2.52 & 4.625 & 5.05 & 4.55 & right\\ 
1.0 + 0.001 & 0.5 & 1.25 & 3.4 & 3.16 & -1.03 & 4.75 & 5.05 & 4.75 & left\\ 
\enddata
\tablecomments{$t_{i}$: time of ignition; 
 $\rho_{\rm{stream}}$: peak density of the stream;
 $\rho_{i}$: density at the ignition point;
 $d_{i}$: distance of ignition point from the impact point. Negative value implies that ignition is on the left of the stream impact point and Positive value implies that ignition is on the right of the stream impact point;
 $H_{i}$: height of ignition point from the center of the WD;
 $d_{tr1}$: distance of the transition region from the center of the WD at t = 0 s;
 $d_{tr2}$: distance of the transition region from the center of the WD at $t_{i}$ \\
 Propagation direction refers to the direction in which the detonation propagates around the entire WD. The detonation can either propagate toward the left, the right or both the left and the right. Starred case (\**): Lower resolution simulation ( cell size $\approx 20$ km). Double Starred case (\***): two detonations ignite one after the other at different points on the domain. But they end up colliding and fizzling out. $d_{tr2}$ is listed as N/A for cases where the final position of the transition region is not obvious because there might have been some prior helium burning before the ignition.}  
\end{deluxetable}



\section{Discussion}  \label{sec:discussion}

Our results primarily show that a mass-transfer stream can ignite a helium detonation on the surface of a WD in two dimensions across a range of carbon-oxygen WD models with helium shell composition profiles that realistically reproduce those left at the end of the prior helium shell burning phase. However, in an expansion upon cases seen in our previous work, the ignition point in some cases is located far from the stream impact point, as shown in Figure \ref{fig:plot4}. According to the D$^6$ model, mass transfer from the companion WD is expected to impact the surface at a small angle away from the line connecting the centers of the two WDs. Figure \ref{fig:plot4}, however, suggests that ignition can occur significantly offset from the impact point, at distances of up to $\sim 10^{9}$ cm. This indicates that ignition is not restricted to the hemisphere facing the companion and can occur anywhere along the equatorial line around the WD surface.

Previous double-detonation studies \citep{Townsley_2019, Boos_2021, Gronow2020, Gronowetal2021} have shown that the secondary detonation typically forms on the opposite side of the WD (in the core), a small distance from the symmetry axis through the helium ignition point, thereby contributing to the asymmetric structure of the ejecta. This means that if helium ignition can occur anywhere around the surface, then the helium ignition location also determines the axis of the asymmetric ejecta of the primary star. The location of the companion, which may or may not explode \citep{Boos2024}, as well as the mild asymmetry of the primary induced by the tidal force of the companion, is therefore not fixed in alignment with respect to the explosion asymmetry. Thus ignition location and companion location may represent two independent tertiary parameters,
with the primary and secondary masses (or total and mass ratio) being the primary and secondary parameters.

Our results further show that all cases ignite a helium detonation. However, (1) the ignition mechanisms differ, and (2) not all cases lead to a propagating detonation.

\begin{enumerate}
    \item In 3 of 4 cases with a thick stream, ignition occurs via a direct mechanism in which the stream impact compresses the surface to sufficiently high temperatures and densities, triggering ignition near the impact point (see Figure \ref{fig:plot9}). The exception is the 1.0 $\,M_\odot$ thick stream case, which ignites far from the impact point. 

    In contrast, 3 of 5 thin stream cases ignite far from the impact point as seen in Figure \ref{fig:plot4}. The ignition mechanisms for the 0.8 $\,M_\odot$ and 0.9 $\,M_\odot$ thin stream cases are illustrated in Figures \ref{fig:plot7} and \ref{fig:plot8} respectively. As discussed in Section \ref{sec:results}, the ignition for these two cases follows the occurrence of a hot spot in a $^{4}$He-rich pocket inside the outer edge of the dense $^{12}$C-rich core. This suggests that ignition can arise in highly disturbed surfaces through nontrivial mechanisms. The 1.0 $\,M_\odot$ and 1.0 + 0.001 $\,M_\odot$ cases, however, ignite in a manner similar to the direct mechanism shown in Figure \ref{fig:plot9}. Additionally, the 1.0 + 0.001 $\,M_\odot$ thin stream with lower density case also ignites via this direct mechanism, contrary to expectations.
    
    Some cases exhibit multiple ignition points, either occurring simultaneously (as in the 1.0 $\,M_\odot$ thick stream and 0.8 $\,M_\odot$ thin stream cases) or at different times (e.g., the 1.0 + 0.001 $\,M_\odot$ thin, lower-density stream case). The implications of multiple helium ignition spots on the double-detonation mechanism has previously been studied by \citet{MollWoosley2013}. In the 1.0 $\,M_\odot$ thick stream case, detonation fails because two simultaneously formed detonations collide and disrupt each other. It is unclear whether this outcome would persist in three dimensions, where detonations are not confined to a plane. In 3D, such interactions may instead strengthen the detonation in other directions, as shown by \citet{MollWoosley2013}, who found that interacting helium detonations can survive and even generate hot spots necessary for the carbon detonation in the carbon–oxygen core.

    \item Not all ignition events lead to a propagating helium detonation. In several cases, the detonation propagates only in one direction along the WD surface. When ignition occurs far from the impact point, propagation often fails in the direction toward the stream. After impact, material rebounds and flows along the surface, predominantly downstream with respect to the incoming stream. This flow can oppose a detonation propagating toward the stream, inhibiting its advancement and resulting in one-sided propagation.

    This behavior is likely a limitation of the two-dimensional geometry. In three dimensions, detonations are expected to propagate in ring-like structures, and a non-propagation in one direction would only imply an asymmetrically propagating detonation. In 2D, however, a one-sided propagation may also reflect local variations in fuel availability, such as reduced helium density or insufficient $^{12}$C abundance.

    \end{enumerate}

Figure \ref{fig:plot5} shows the height of the center of the transition region at the ignition point from the WD center, at $t = 0$ s (labeled as the initial transition region) and at $t = t_{i}$ (labeled as final transition region) along with the distance of the ignition point from the center of the WD. The transition region is a thick layer where the $^{12}$C and $^{16}$O abundances gradually decrease while $^{4}$He abundance gradually increases. We define the location of the transition region as the point where the $^{4}$He and $^{12}$C abundances are equal. As the stream deposits material onto the surface, the accreted material pushes the transition region deeper into the WD and the helium ignition occurs near the location of the final transition region.    

\citet{Shen_2024} showed that the 1.0 $\,M_\odot$ WD does not support a propagating detonation, even at higher resolutions, because the density at the transition region is insufficient for the detonation to propagate in it. In our simulations, however, both 1.0 $\,M_\odot$ cases ignite only after $\sim 4$ s by which time a significant amount of material has been accreted to the WD surface, compressing the transition region to higher densities. Although both cases ignite, only the thin stream case sustains a propagating detonation. Figure \ref{fig:plot6} compares the densities at the initial and final transition regions and shows that the ignition occurs at higher densities for all cases compared to the undisturbed shells in \citet{Shen_2024} since all the ignition points lie close to the final transition region. 

From Table \ref{tab:deluxesplit}, the 1.0 + 0.001 $\,M_\odot$ cases ignite earlier and via the direct mechanism, whereas the 1.0 $\,M_\odot$ cases ignite later after significant surface disruption. This suggests that a certain amount of accreted material may be necessary to achieve ignition conditions. Further supporting this, the thin + lower density stream case ignites at $t = 3.4$ s, despite having a lower stream mass impacting the surface and thus expected to ignite later. This indicates that pre-existing accreted material may facilitate ignition.

The 0.9 $\,M_\odot$ thin stream case is particularly interesting because after the first wrap-around, the detonation continues to propagate in the lower density accreted helium layer, suggesting that detonations may also propagate in such extended surface layers.

There are several limitations to our simulations. First, our simulations are performed in two dimensions. In three dimensions, the ignition mechanism and detonation morphology may differ due to the additional spatial degree of freedom and the more realistic geometry of the accretion stream. In 2D, the flow of accreted material on the WD surface is constrained to a slice which could affect the ignition mechanism, and the ignition process may therefore differ in 3D. For this reason, although the present study demonstrates the feasibility of helium ignition under the adopted stream conditions, the detailed ignition mechanism remains to be established in 3D. We will address this in follow-up work.

Second, the simulated surface domain is smaller than the full stellar circumference. For example, the circumferences of the 0.8, 0.9, 1.0, and 1.0 + 0.001 $\,M_\odot$ WDs are $\sim 4.30 \times 10^{9}$ cm, $3.83 \times 10^{9}$ cm, $3.36 \times 10^{9}$ cm, and $3.37 \times 10^{9}$ cm, respectively, whereas our domain length is $2.00 \times 10^{9}$ cm. As a result, detonations traverse a shorter distance than in reality, which may affect ignition location and timing. We adopt this reduced domain to limit computational cost, noting that \citet{Rajavel2025} found that domain size primarily affects ignition time rather than whether ignition occurs. Third, it is difficult to estimate what the stream density and width should be for a particular mass ratio at a particular stage of the merger without a detailed treatment of the evolution of mass transfer in the binary system. Therefore, we instead used a stream density and width that could potentially ignite the surface, as shown in \citet{Rajavel2025}, rather than parameters we would expect for a given binary system. Consequently, the adopted stream parameters represent an idealization, and it remains to be seen how the ignition mechanism would change for stream conditions arising from realistic binary evolution.

Importantly, our results demonstrate, for the first time, that stream-initiated helium detonations can be produced using realistic WD profiles with thin helium shells and smooth core–shell transition regions. This is important because, once the detonation has propagated around the WD, our models do not show significant amounts of high mass elements (atomic number $> 21$) being produced, and overall, none of our simulations show any elements heavier than $^{48}$Cr with mass fractions above 0.01 being produced, with the heaviest elements that are produced occurring only in very thin layers. This is expected to be consistent with observations of Type Ia supernovae, which do not show high-mass elements at high velocities in the ejecta. 

Although the present study demonstrates the feasibility of helium ignition under the adopted stream conditions, the detailed ignition mechanism remains to be established in future 3D work.

\section{Summary}  \label{sec:conclusion}


We perform two-dimensional hydrodynamic simulations of a mass-transfer stream impacting a plane-parallel approximation of a CO WD surface using the FLASH code, following the setup of \citet{Rajavel2025}. We adopt realistic one-dimensional WD profiles from MESA \citep{Shen_2024} for 0.8, 0.9, 1.0, and 1.0 + 0.001 $\,M_\odot$ WDs and map these profiles into our simulations. The resulting profiles are those produced by the natural process of the
extinguishing of shell burning at the end of stellar evolution. For each WD mass, we explore two stream thicknesses to investigate whether helium detonations can be robustly ignited under realistic conditions and to characterize the resulting ignition and propagation behavior.

We find that all cases ignite a helium detonation, and all but one successfully develop a propagating surface detonation. A key result is that ignition is not confined to the stream impact point and can occur at a wide range of locations around the WD surface. In all cases, ignition occurs near the core–shell transition region, highlighting the importance of using realistic WD profiles for accurately capturing ignition conditions and detonation propagation.

Our simulations also show that nucleosynthesis is dominated by intermediate-mass elements, primarily $^{28}$Si and $^{32}$S, with only trace production of heavier elements. No elements beyond $^{48}$Cr are synthesized in significant quantities. This further highlights the importance of realistic initial profiles in reproducing observational constraints of Type Ia supernovae.

Additionally, we find that ignition can occur in highly disturbed surface layers and proceeds through nontrivial mechanisms that depend on local conditions. Although the present study demonstrates the feasibility of helium ignition under the adopted stream conditions, in future work we will extend these simulations to three dimensions to better capture the geometry of the accretion stream, the WD surface and the resulting detonation structure.

Animated plots of the log density, temperature and log mass fraction evolution for all cases discussed in this paper are available at the following link:
https://doi.org/10.5281/zenodo.21384193
\begin{acknowledgments}

The software used in this work was developed in part by the DOE NNSA- and DOE Office of Science-supported Flash Center for Computational Science at the University of Chicago and the University of Rochester. Resources supporting this work were provided by the NASA High-End Computing (HEC) Program through the NASA Advanced Supercomputing (NAS) Division at Ames Research Center. N.R. received support from NASA FINESST under award No. 80NSSC23K1438. D.M.T and N.R. also received support from the National Science Foundation under grant No. 2307442. K.J.S was supported by NSF grant AST 2508988 and NASA/ESA Hubble Space Telescope program No. 17441.

\end{acknowledgments}

\software{Flash 4.6 \citep{Fryxell, DUBEY2009512, Dubey2013, Dubey2014},  
          MESA \citep{Paxton2011, Paxton2013, Paxton2015, Paxton2018, Paxton2019, Jermyn2023}, 
          yt \citep{yt}
         }



\bibliography{sample701}{}
\bibliographystyle{aasjournalv7}

\end{document}